\documentclass[preprint, journal]{vgtc}            

\onlineid{0}

\preprinttext{To be published by IEEE Xplore.}

\vgtccategory{Research}

\vgtcpapertype{Position Paper}

\title{Complexity as Design Material \\
\large Position Paper}

\author{%
  \authororcid{Florian Windhager}{0000-0002-5170-2243},
  \authororcid{Alfie Abdul-Rahman}{0000-0002-6257-876X}, \authororcid{Mark-Jan Bludau}{0000-0001-6300-8833}, \authororcid{Nicole Hengesbach}{0000-0001-7918-6343}, \authororcid{Houda Lamqaddam}{0000-0001-6738-1934}, \\
  \authororcid{Isabel Meirelles}{0000-0001-8111-6002}, \authororcid{Bettina Speckmann}{0000-0002-8514-7858}, and 
  \authororcid{Michael Correll}{0000-0001-7902-3907}  
}

\authorfooter{
  \item
  Florian Windhager is with University for Continuing Education Krems.
  E-mail: florian.windhager@donau-uni.ac.at
  \item
  Alfie Abduhl-Rahman is with King's College London.\\
  E-mail: alfie.abdulrahman@kcl.ac.uk.
  \item Mark-Jan Bludau is with University of Applied Sciences Potsdam.\\
  E-mail: mark-jan.bludau@fh-potsdam.de.
  \item
  Nicole Hengesbach is with Warwick University.\\
  E-mail: n.hengesbach@warwick.ac.uk
  \item
  Houda Lamqaddam is with University of Amsterdam.\\
  E-mail: h.lamqaddam@uva.nl.
  \item Isabell Meirelles is with OCAD University Toronto.\\
  E-mail: imeirelles@ocadu.ca.
  \item
  Bettina Speckmann is with Eindhoven University of Technology.\\
  E-mail: b.speckmann@tue.nl.
  \item
  Michael Corell is with Northeastern University.
  E-mail: m.correll@northeastern.edu.
}

\abstract{%
  Complexity is often seen as a inherent negative in information design, with the job of the designer being to reduce or eliminate complexity, and with principles like Tufte's ``data-ink ratio'' or ``chartjunk'' to operationalize minimalism and simplicity in visualizations.
  However, in this position paper, we call for a more expansive view of complexity as a \textit{design material}, like color or texture or shape: an element of information design that can be used in many ways, many of which are beneficial to the goals of using data to understand the world around us.
  We describe complexity as a phenomenon that occurs not just in visual design but in every aspect of the sensemaking process, from data collection to interpretation.
For each of these stages, we present examples of ways that these various forms of complexity can be used (or abused) in visualization design.
  We ultimately call on the visualization community to build a more nuanced view of complexity, to look for places to usefully integrate complexity in multiple stages of the design process, and, even when the goal is to reduce complexity, to look for the non-visual forms of complexity that may have otherwise been overlooked.
}


\teaser{
  \centering
  \includegraphics[width=\linewidth, alt={Axes of complexity transformation in visualization design.}]{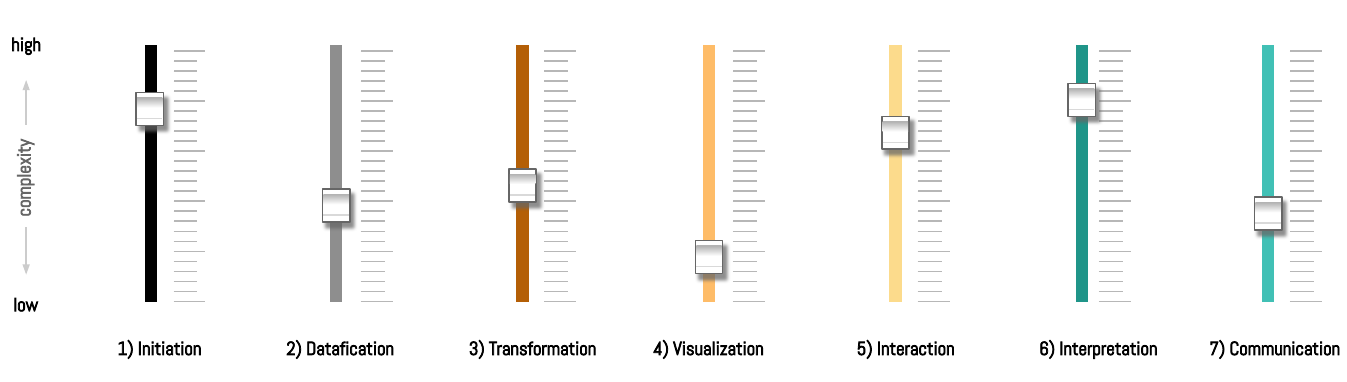}
  \caption{%
  	Axes of complexity and complexity transformation in visualization design, bridging from project initiation complexity to the complexity of interpretation and communication activities, using the metaphor of a mixing board. A designer might strategically employ higher or lower levels of complexity across these axes to achieve a desired effect. Likewise, changes to one type of complexity \textit{shift} complexity to other parts of the pipeline.
   }
  \label{fig:teaser}
}

\graphicspath{{figs/}{figures/}{pictures/}{images/}{./}} 

\usepackage{tabu}                      
\usepackage{booktabs}                  
\usepackage{lipsum}                    
\usepackage{mwe}                       

\usepackage{mathptmx}                  

\begin{document}

\definecolor{initiation}{HTML}{000000}
\definecolor{datafication}{HTML}{8e8e8e}
\definecolor{transformation}{HTML}{b45f06}
\definecolor{visualization}{HTML}{e69636}
\definecolor{interaction}{HTML}{eabe54}
\definecolor{interpretation}{HTML}{1d8e8f}
\definecolor{communication}{HTML}{41c0b5}



\firstsection{Introduction}
\label{introduction}
\maketitle

Much ink has been spilled and experimental data collected in service of what is perceived as a key conflict in data visualization research and practice: the role of complexity. On one side, a ``minimalist''~\cite{inbar2007minimalism} view of data visualization design towards reducing complexity as much as possible, influenced by notions from Tufte such as ``chartjunk'' and the ``data-ink ratio''~\cite{tufte2014visual}. On the other side, a view touting the benefits of adornment and ornamentation on grounds of memorability~\cite{borkin2013makes}, pedagogy, or a host of other desiderata~\cite{bertini2020shouldn}. In short, to what extent is the goal of data visualization to reduce complexity?

We posit that certain kinds of complexity, especially when thinking of complexity \textit{beyond the visual}, is rarely, if ever, actually reduced, but merely \textit{moved} to other parts of the process of visualization creation or interpretation. We draw inspiration here from maxims like Tesler's Law that states that ``every application has an inherent amount of irreducible complexity. The only question is who will have to deal with it, the user or the developer''~\cite{saffer2010designing}. In line with the adage ``as simple as possible, but not simpler'' we consider an adapted \textit{law of requisite complexity} \cite{boisot2011complexity} as a noteworthy addition to reductionist visualization design guidelines: in the face of complex things, useful representations have to actually \textit{retain} a sufficient degree of complexity, to inform related controlling, governing and problem solving efforts.
Lastly, we hold that complexity is not an intrinsic deficiency, or even an overall negative component with occasional counter-intuitive ``beneficial difficulties''~\cite{hullman2011benefitting}, but that complexity is (beyond being an inescapable and often necessary facet of human experience~\cite{norman2016living}) frequently a \textit{useful} tool for accomplishing important design goals.

We therefore propose that neither the thesis of data visualization minimalism nor the antithesis of data visualization adornment capture the phenomenon of interest or the language in which we should be articulating design principles. We instead propose a synthesis: \textit{complexity as a \textbf{design material} that can be strategically employed by designers at all stages of the design pipeline}. Just as a designer can make judicious use (or non-use) of color, interactivity, or animation to accomplish design goals in data visualization, so too can a designer thoughtfully employ (or even ``dance with''~\cite{garcia2014enactive}) complexity. To continue the analogy with color, when, for example, Stone proposes that chart designers seeking to employ color first ``get it right in black and white''~\cite{stone2010get}, this is not to divide the community into pro-color and anti-color camps, but to call for intentionality and strategy in the use of color as a design material. Similarly, we hold that designers should be mindful and strategic in how they employ complexity, rather than a knee-jerk assumption that complexity can (or should) be unilaterally minimized in a funnel-like process of reduction from data to chart to insight. 

To guide our discussion of complexity's role in the design process, and to focus on forms of complexity other than merely the visual, we propose a set of \textit{axes of complexity} (\autoref{fig:teaser}) that represent complexities at all stages of the design process, and show how complexity changes form, purpose, and role as it moves along these axes. We focus especially on aspects of complexity that have been given short shrift by the over-emphasis on the pros and cons of strictly \textit{visual} notions of visualization complexity. We suggest that there are many existing productive or fruitful designs of visualizations that employ strategies other than reducing complexity as much as possible, as quickly as possible, as prescribed by many of the standard visualization pipelines. 

We also suggest that often what is perceived as a \textit{reduction} in complexity is in fact a \textit{shifting} of this complexity to other points along the design and sensemaking process. For instance, while a technique like principal component analysis could be used to reduce a high-dimensional dataset down to a two-dimensional scatterplot---a seeming reduction in data complexity---the resulting axes of this plot are now esoteric and difficult to interpret~\cite{gleicher2013explainers}: complexity has not been removed, but shifted to a complexity in \textit{interpretation}. 

Where and how complexity is shifted during visualization design should be the result of a careful balancing act and negotiation between designers, researchers, audiences, and many other stakeholders, and not the result of standardized complexity reduction and control.
We hope that considering the movement and translation of complexity along our proposed axes might reveal places where novel design interventions or design goals could be realized, and places where we lack current guidelines or empirical guidance from the visualization literature.

The rest of the paper is organized as follows. We first employ on thinking about complexity both within and without the field of visualization to lay out existing notions of complexity (\S\ref{sec:relatedwork}). We aim to integrate, enrich, and contextualize these existing notions with a process-oriented model of complexity along seven axes (\S\ref{sec:axesofcomplexity}). We conclude with a call to action for the visualization community to adopt a more nuanced and multivalent view of complexity in visualization design, and in particular to draw more attention to the forms of complexity beyond the visual that are often not well-discussed in existing visualization empiricism or pedagogy (\S\ref{sec:discussion}).

\begin{figure*} [th!]
    \includegraphics[width=\textwidth, alt={Recurring definition criteria of ``complex'' subject matters (left) and phenomena to which this quality is frequently ascribed.}]{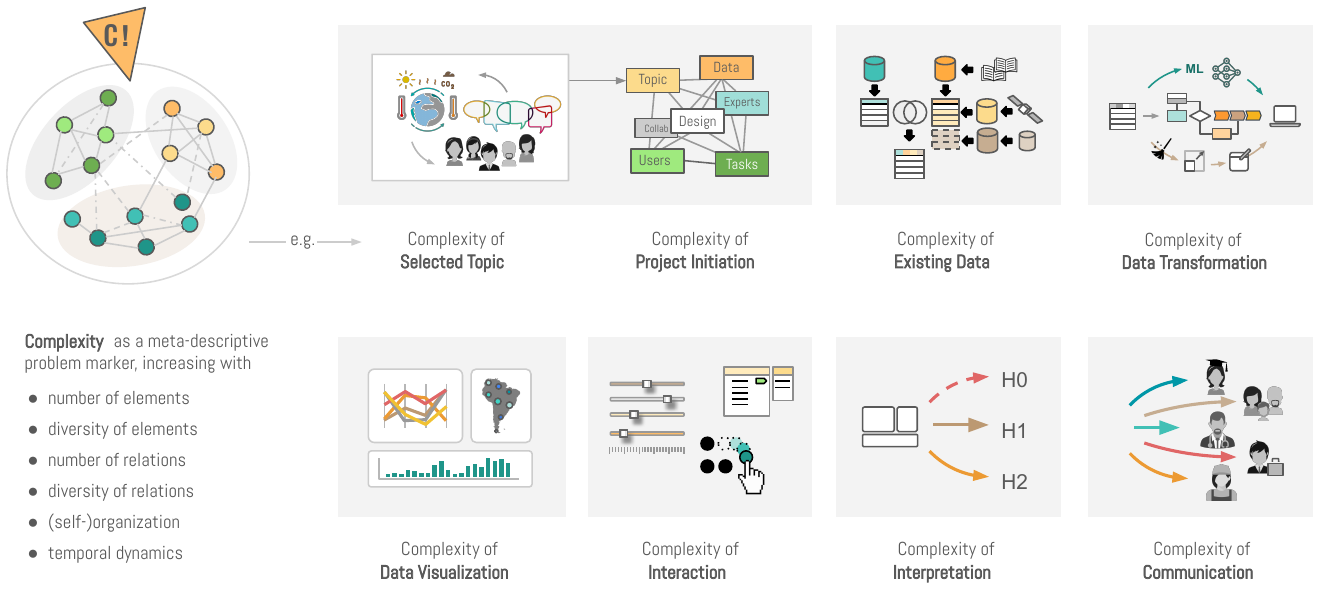}
  \caption{Recurring definition criteria of ``complex'' subject matters (left) and application of this definition to different aspects of visualization practice.
  }
  \label{fig:def-cases}
\end{figure*}

\section{Related Work}
\label{sec:relatedwork}
In this section, we briefly overview complexity as a general concept 
and then provide an overview of existing notions of complexity in visualization.

\subsection {Complexity in General}

``Complexity'' as a concept is used to characterize a wide range of topics and phenomena in a large number of knowledge domains, including physics \cite{bennett1990define}, biology~\cite{mazzocchi2008complexity}, psychology~\cite{woods1988coping}, sociology~\cite{castellani2009sociology}, economics~\cite{arthur2021foundations}, management \cite{olson2001facilitating}, technology~\cite{kallinikos2005order}, and in cross-domain discourses such as systems and complexity science~\cite{flood2013dealing, phelan2001complexity}. In common parlance, the adjective ``complex'' is closely tied to---and often mostly synonymous with---``complicated'' or ``intricate'' (although see Norman~\cite{norman2016living} for a theoretical distinction between ``complex'' as a property of user interpretation and ``complicated'' as a property of the exterior world). In contrast, many scholars have argued to reserve the use of ``complex'' to phenomena with a very large numbers of \textbf{differentiated elements and interactions}, frequently including increased  levels of \textbf{descriptive ``unknowns'' and ``unknowables''}, which leads to a substantial \textbf{decrease of external comprehensibility and controllability}~\cite{glouberman2002complicated, kinni2017critical, poli2013note}, and the need for phenomena to \textbf{``self-organize'' and build up order internally}, from structural hierarchies to adaptive behavioral patterns.

For the rest of this paper, we use the term ``complexity'' with this whole spectrum of meaning in mind---from a general concept to system-specific definitions, rather than focus on one particular definition divorced from a context of use. Defining complexity in a rigorous fashion \textit{aside from certain fields of study} has been called an ``intractable problem''~\cite[p.52]{taborsky2014complexity}. This has sparked controversies about the scientific value of the concept in general \cite{teixeira2020thinking}, but its ever-increasing use across all fields of societal discourse rather proves the value of the term as a transdisciplinary ``problem marker''.  

In approximation to information-theoretical definitions, this problem marker often refers to things or processes, which evade simple descriptions and attempts of control---relative to an established level of descriptive and cybernetic (i.e. management or control-oriented) resources. More specifically, ``complex'' is used to flag subject matters, which consist of a \textit{large number of differentiated elements}, which interact in varied and often unpredictable ways, due to \textit{numerous and diverse intra-system and system-environment relations and interactions}. In addition, a diachronic perspective (i.e. observation over time) often describes \textit{non-trivial and unpredictable system behaviors}. 
Systems-theoretical perspectives often reflect on the ``emergent'' qualities of complex systems, which are said to create novel, ``super-summative'' characteristics and behaviors, which are non-reducible to an atomistic understanding of their parts, as complexity often leads to self-organizing and adaptive behavior \cite{sawyer2005social, mann2012systems, heylighen2001science}.

According to information-theoretical definitions, complexity largely corresponds to the length of system descriptions needed to successfully model (i.e., describe, calculate, explain, compute, comprehend, communicate) and govern a system and its behavior \cite{hale2016information}. Thinking and learning about complex things (i.e. building up viable mental models) and training how to handle them takes time---a burden which can be shared by dividing knowledge and labor. Many systems scientists diagnose a general trend in the evolution of societies and other systems towards an increase of internal complexity \cite{kauffman1993origins, mcneill2015human}. Given complex and adversarial environments, systems with more knowledge and response options gain a competitive evolutionary edge. The \textit{``law of requisite variety''} \cite{klir1991requisite} states that ``The larger the variety of actions available to a control system, the larger the variety of perturbations it is able to compensate'' or (only) ``complexity absorbs complexity''. By raising their socio-technical complexity (e.g., by dividing roles, knowledge and labor between individuals and organizations as task-specific problem-solvers) cultures and systems in general are able to address increasing amounts of problems\cite{richerson2008not, wenke1981explaining, rescher2020complexity}. However, this collective strategy also increases the complexity challenges for individual observers, whose natural perceptual and cognitive resources remain limited.

\subsection{Complexity in Visualization}

The outlined complexity of natural, cultural, technological, and semiotic environments provides the standard argument for the \textit{cognition-enhancing technologies} of modern times, including the tools of visualization \cite{hayles2012we,liu2008distributed}. Given the raw number of complex subject matters---and their amassed descriptions in libraries, archives, and interlinking data collections, ``second order technologies'' become key. These technologies augment and amplify human cognition in face of complex (i.e. ``massive, dynamic, ambiguous, and often conflicting'' \cite[p.10]{thomas2006visual}) data.

Visualization thus aims to translate data complexity into forms that reduce the cognitive load of sensemaking \cite{paas2010cognitive}, mediating between topic and data complexity on the one hand and limited cognitive bandwidth on the other ~\cite{arias2012cognitive}. At first glance, this objective favors aesthetics of simplicity, efficiency and reduction to the essence. Just as a map cannot and should not mirror the territory, neither can or should a diagram represent the full complexity of the underlying topic. Thus, a \textit{complexity-reductionist agenda} undergirds many visualization design guidelines: visualizations should move us from data to insights, help to separate the signal from the noise, and put related evidence into a proper visual perspective. Above all else, visualizations should show the data, steer clear from all other distractions, and generally generate more (insights) with less (ink/pixels) \cite{tufte2014visual}. 

According to these tenets, visualization pipelines serve as funnels for data complexity which minimize the ``data-to-cognitive-work ratio'': The time and effort needed to process a given dataset alphanumerically is significantly reduced by visual encodings (if in line with rules of ``graphical excellence'')---so that cognitive activities (including visual perception, mental modelling, reasoning and communication) can happen for a fraction of the original processing costs. If individual views cannot exhaust a dataset, facets of data complexity will not be visually superimposed, but distributed over \textit{multiple} minimalist views. Tellingly, this should happen in line with the \textit{rule of parsimony} that recommends to keep the number of views to a minimum \cite[p.115]{baldonado2000guidelines}. 

While these minimalist design aesthetics have a pervasive and longitudinal impact on the visualization field, there are empirical and theoretical objections to the orthodox application of minimalism in visualization. Concepts such as ``data-ink-ratio'' or ``chart junk'' have attracted the scrutiny of visualization researchers, who have collected both empirical evidence against an overly-narrow application of minimalist strategies in visualization design as well as counter-examples or counter-framings~\cite{inbar2007minimalism, bateman2010useful,chen2010information,hill2018minimalism,hullman2011benefitting, li2014chart,behrisch2018quality,parsons2020data, ajani2021declutter,akbaba2021manifesto, hill2016visualizing}. 

Arguably, this debate around minimalism (see also \S\ref{sec:visualization}.4) has largely dominated the discussion of complexity in the visualization field. However, the state of the debate is still quite confrontational and centers on visual complexity, while the multi-faceted concept of complexity we propose offers the chance to synthesize discussions about consequential translations and trade-offs across multiple dimensions and steps of the design process, and to do so in a more balanced and systemic fashion. Thought of as design material, complexity can be understood as a resource, rather than a problem marker only, that can be used to achieve different ends. Figure~\ref{fig:def-cases} (left) draws together definitional facets of complexity which we find in the literature (i.e., many elements or options, many relations or interactions, diversity of elements and relations, relevance of time) and illustrates how they can help to characterize the distributed complexity scenarios in visualization (right). As a design material, complexity can be used in many places: sparingly used, say, in visual design, but lavished on interaction design or in data preparation. In the following sections, we collect and connect related observations and arguments on either side of the focus area of \textit{visual} complexity, and thus connect to other major decision scenarios for the design and reading of visualizations---what we term ``axes'' of complexity along the visualization design pipeline.

\section{Axes of Complexity}
\label{sec:axesofcomplexity}

In this section, we focus on areas where complexity can arise in the process of visualization, with a particular focus on facets other than the complexity of a visualization \textit{per se}. We structure these complexities along a set of axes capturing key decision points in the visualization design process, from selecting topics and hypotheses of interest to the communication of any derived insights to stakeholders. (The font color of the subsections correspond to the axes introduced in Figure \ref{fig:teaser}.)

The concept of ``axes'' serves as a heuristic device to mark decision points of complexity regulation or transformation along the sensemaking pipeline (e.g., \cite{mcnutt2020surfacing, haghighatkhah2022characterizing}).
While we discuss our axes in a separated fashion (acting like sliders of a mixer console, see \autoref{fig:teaser}), in practice, these processes have blurred areas of responsibility, overlap each other, or are tightly coupled (e.g. interaction may lead to direct data transformations and encoding changes). Rather than present a complete and compartmentalized view of sensemaking using visualization, it is the general aim of the following sections to highlight sources of complexity that might otherwise be missed in the familiar discussion of \textit{visualization complexity} (focusing on the visual) and to bring connected consequences of any complexity decision into a sharper focus.

\textcolor{initiation}{\subsection{Initiation}}  
\label{initiation}

\noindent Initiating visualization projects entails numerous decisions about the overall complexity of the ensuing setup, including deliberations on all the following phases and axes. At the outset, there is a \textit{topic} of interest (or one is selected), for which knowledge has to be built up, or around which collaborators with extensive prior domain knowledge exist~\cite{sedlmair2012design}. This meeting point between domain knowledge and visualization process is also a site of friction and heightened complexity. At this frontier, complexity stems from two complementary aspects: 

\noindent (1a) \textbf{Topic complexity}: No matter the discipline, the original phenomena, topics, or study materials can be of varying levels of complexity depending on the nature of the material as well as the goals addressed. As an example, we look at the phenomenon of analyzing and communicating climate change data to policy makers or the wider public \cite{schuster2024being, metze2020visualization, windhager2019inconvenient}, which might involve forecasting of many interconnected factors like temperature, sea level rise, precipitation, etc. The studied phenomenon in this case rests on a multivariate interaction of several parameters over time, supported by complex mathematical and statistical analysis. Narrowing the sorts of questions involved (say, from climate change as a gestalt phenomena to merely one aspect, like sea level rise) may still rest, at heart, on complex and interconnected \textit{phenomena}, but reduces the scope of the topic at hand and has a large impact on the downstream complexity of any visualization project.

\noindent (1b) \textbf{Descriptive complexity}: Topic complexity arguably translates into complexity of topic descriptions and representations---whether of texts, equations, models, or algorithms, which can be approximated with measures such as linguistic or algorithmic complexity, and secondarily with regard to depth, variety and history of related knowledge, discourse, and controversies (e.g., for climate change, see \cite{grundmann2007climate, porter2018organizing}). As the complexity of topics arguably cannot be measured directly, we treat descriptive complexity as a proxy for topic complexity.

\noindent (2) \textbf{Epistemic differences} between the domain discipline and visualization research: The distance in epistemic frameworks between the application domain and visualization is a compounding factor for topic complexity. In computational disciplines, the complexity of the discipline is largely due to what we describe above as inherent to the data or tasks. It can then be relatively smoothly translated into data, and transformed into a design and interaction design task. Rather than a straightforward translation of phenomenon or topic into data, many disciplinary collaborations require a broader gap to be bridged in order to align the studied objects with visualization-specific paradigms. In disciplines within the humanities, for example, scholarship is interpretative, qualitative, often focused on a small dataset or singular object of study. The inherent complexity of the task and theoretical framework is then supplemented by the additional effort to bridge the epistemic and methodological differences \cite{bradley2018visualization,hinrichs2017risk}.

\smallskip\noindent
\textbf{Uses and abuses:} These sources of complexity suggest that the interaction between domain phenomena and visualization inherently makes complexity a core element of the design process from the very start of the visualization process. The amount of attention that is initially given to a topic's epistemic, descriptive and discursive complexity is a critical factor, deciding the success of a project (e.g., whether it will be considered to be a legitimate contribution, or rather a ``trojan horse'', bringing undue computational assumptions and simplifications to the field \cite{drucker2011humanities}). Visualization design allows for the reduction of topic complexity for instance by abstracting or re-scoping tasks, focusing on a case study, or otherwise performing ``data counseling''~\cite{fisher2017making} during the process of operationalization. However, the insights and interpretations that can be taken from data and its visualization in relation to the real-world phenomenon depend on this translation. If the phenomenon becomes oversimplified through datafication or visualization, insights will remain limited based on the (semantic or causal) gap between what is represented and how it is represented \cite{lamqaddam2020introducing, offenhuber2019data}. 
\\

\textcolor{datafication}{\subsection{Datafication}}
\label{datafication}

\noindent Datafication is the process of transforming aspects of a real-world phenomenon or topic into data that can be processed computationally for different purposes, and is an active process of shaping rather than a mere passive collection (indeed, Drucker~\cite{drucker2015graphical} uses the term ``capta'' over the usual ``data'' to precisely denote this active and teleological process). Datafication includes activities of a) native data creation and collection from observations or digital sensors, as well as b) digitization practices which convert analog content or representations of a subject matter---such as books, films, photographs---into digital information.  In our context, datafication also includes c) the selection and compilation of given datasets about a topic, to either work with them directly or to prepare them for further transformation. While many of these processes are fairly standardized, they concatenate, aggregate and introduce a substantial number of (ontological, epistemological, methodological, practical) design choices and thus are acutely non-trivial procedures, which can be done more or less comprehensively \cite{southerton2022datafication, flensburg2023datafication}. 

\textit{Datasets for any object of study thus can be more or less complex}: Simply put, low complexity datasets appear as spreadsheets with small numbers of rows, columns, facets, or data types, while high complexity data sources aggregate multiple data sources into large numbers of rows, columns, facets, data types, and relations---with fundamental implications for all (computational, analytical, or interpretational) downstream activities. Descriptions of datasets, such as datasheets~\cite{gebru2021datasheets}, can therefore contain a myriad of factors, from sourcing to metadata to an assessment of potential biases and limitations of the data that have been collected.

When thinking about complexity as ``design material'', datafication is a major decision point, even though it's often overlooked for two reasons: (1) As visualization researchers often work with pre-existing data sets and sources, they tend to refer back to domain experts for decisions on both topic and datafication complexity, in order to jump straight to transformation activities, and (2) Visualization experts are used to starting from high data complexity as a \textit{fait accompli}: data is mostly complex, that's why the visualization design expertise is needed to distill it into simpler, more accessible and understandable formats. That is, there is a selection bias in visualization research, where we often focus on complex (but exotic) ``zebra'' problems while giving shorter shrift to more quotidian ``horse'' problems~\cite{correll2020we} that can be solved by existing methods and techniques.

\smallskip
\noindent\textbf{Uses and abuses:} Given this dominant notion of pre-existing data complexity, related design questions and complications have been discussed on various levels:

\noindent (1) Datasets can be complex, but full of errors, problems, and issues which leads to the next phase of \textit{transformation} activities.

\noindent (2) Datasets can be correct on their faces, but represent just one perspective or descriptive option amongst many---especially for topics with a long and complex, controversial discursive history.

\noindent (3) Datasets can be constructed to deliberately hide, distort, mislead with respect to a topic. Even without intentional malicious decisions to bias, there is no objectively created dataset that represents a ``view from nowhere''~\cite{haraway2013situated}: the situatedness that we bring to knowledge also is reflected in how we construct datasets.

\noindent (4) Even corrected and enriched by further transformations, datasets can be complex, but not complex enough to e.g. enable or support meaningful analysis and interpretation of a subject matter. This is a challenge extensively discussed in the humanities, whose language-based inquiries allow for all kinds of quantitative expressions, while largely focusing on qualitative assessments, interpretations and arguments. Datafication and computational methods with their focus on quantification thus significantly reduce and limit the existing means of ``mixed methods'' modeling, sensemaking, and reasoning \cite{drucker2020visualization}.\footnote{See also ``dataism'' as modern-day ideology ``rooted in a belief in the capacity of data to represent social life, sometimes better or more objectively than pre-digital (human) interpretations'' \cite{hintz2018digital}.} Without due attention to these datafication-related questions, no downstream means or efforts (including visualization interpretation and communication) will be able to correct related design choices and challenges. However, critical transparency and design strategies \cite{dork2013critical} together with mixed-methods project designs can help to counterbalance detrimental datafication consequences.

\textcolor{transformation}{\subsection{Transformation}}
\label{transformation}

\noindent After a topic and data source has been selected, there is almost always still remaining work before visualization can begin. The process of acquiring, preparing, cleaning, federating, verifying, and otherwise building the data set to be visualized contains many complex steps involving diverse sets of expertise~\cite{alspaugh2018futzing,bartram2021untidy,kandel2011research,muller2019data} that, in repeated surveys of work practices of data scientists, dwarf the time and effort spent on actual visualization and analysis~\cite{anaconda}. Yet, despite the importance and potential depth of these data transformation and preparation steps, relatively few tools exist for visualizing data provenance~\cite{bors2019capturing,callahan2006vistrails,cutler2020trrack}, or the impact of different potential analytical ``paths'' on the subsequent conclusions generated from the data ~\cite{dragicevic2019increasing,sarma2024milliways,pu2018garden}.

The creation of models and other algorithmic ways of enriching data can also be thought of as a form of transformation, with all of the requisite complexities that come from modeling. We note that the complexity of the models \textit{per se} is of interest for this particular axis, rather than the outputs of the models. For instance, dimensionality reduction methods would, from certain perspectives, appear to be a reduction in data complexity (going from many dimensions down to only a few). However, the myriad techniques for dimensionality reduction have varying levels of sophistication, fidelity of the collected data from the datafication step, and many resulting practical interpretations~\cite{espadoto2019toward}. We argue that eventually the price for this reduction is paid later on in the sensemaking process (for instance, in complexity of interpretation). Both modeling and more traditional data preparation steps can interact in unexpected ways that impact downstream complexity as well. For instance, Crisan \& Correll~\cite{crisan2021user} note that (often invisible or comparatively minor) decisions around data cleaning and text processing often had more significant impacts on resulting visualizations of topic models than the hyper-parameters of the actual topic modeling algorithms used.

\smallskip
\noindent\textbf{Uses and abuses:} There are many places where surfacing the provenance of a visualization, or the analytical steps taken to arise at a final data set, can be beneficial, despite introducing considerable complexity in the design process. The transparency, reliability, and credibility of data-driven insights is dependent on truthfully communicating not just results but also methods. Prominent scandals in the ``replication crisis'' and, in some cases, explicit scientific fraud, have been driven by the fact that there are many possible ways to subtly or evenly blatantly manipulate data (and so conclusions) in ways that can be difficult to detect even with direct inspection~\cite{lange2023ferret}. Since data cleaning and transformation also involves many subjective and potentially tendentious choices, it can also be used to discount the credibility of otherwise authoritative datasets~\cite{lee2021viral}.

\textcolor{visualization}{\subsection{Visualization}}
\label{sec:visualization}

\noindent Visualization complexity is arguably the core around which the complexity debate in visualization research has crystallized: If data complexity creates issues for sensemaking, visualizations provide a solution by distilling it into simpler, more accessible and understandable formats. But what is the right amount or level of visual simplicity?

Given a limited amount of effective visual variables and screen space on the one side, and a limited perceptual and interpretational bandwith on the other side \cite{paas2010cognitive}, ``minimalist'' guidelines aim for maximum reduction of structural and formal chart complexity, influenced by notions from Tufte such as ``data-ink ratio'' or ``chartjunk'' ~\cite{tufte2014visual,inbar2007minimalism}. The driving notion has been likened to the modernist pursuit of the ``anti-sublime'': ``If Romantic artists thought of certain phenomena and effects as un-represantable, as something which goes beyond the limits of human senses
and reason, data visualization artists aim at precisely the opposite: to map such phenomena into a representation whose scale is comparable to the scales of human perception and cognition.''\cite{manovich2002data}. If one visualization is not enough, representational complexity should be distributed across the smallest number of multiple views \cite{baldonado2000guidelines}. 

Similar to meta- or postmodernist reflections in art history, a growing body of work built on these minimalist principles but challenged the all too rigid and orthodox applications, emphasizing the many `paradoxical' or plainly beneficial exceptions, where complexity-increasing elements such as adornment and ornamentation strengthen intended effects such as memorability~\cite{borkin2013makes}, pedagogical efficiency, aesthetics, playfulness, joy of use, and a whole range of other desiderata~\cite{bertini2020shouldn, hill2018minimalism,hullman2011benefitting, li2014chart,parsons2020data, ajani2021declutter,akbaba2021manifesto, hill2016visualizing}.

Like other forms of representation (e.g., oral vs. written language), the visual representation and remediation of data necessarily involves abstraction and reduction processes, be that of dimensions in the data or the level of properties, such as when we aggregate or summarize. The outlined discussion thus revolves around the omnipresent danger of eliminating useful complexity, such as the isolation of entities from context, the removal of relevant interdependencies, and the reduction of nuance. During each step of reducing or eliminating data complexity, assumptions are made to achieve a productive level of abstraction. These assumptions must be presented to the target audiences to ensure information is preserved and accurate, and semiotic complexity for the interpretation of a chart is maintained  e.g. by annotations, legends, and captions. As for appropriate design choices, it is relevant to consider comprehensively whether visually preserving the phenomenon's complexity is meaningful or beneficial, despite raising difficulties \cite{hullman2011benefitting}. 

\smallskip
\noindent\textbf{Uses and abuses:} What does visual complexity do to visualization readers and users? While specific reactions depend on context of users and use, the general assumption is that it \textit{raises costs associated with visualization interpretation and communication}---which can be detrimental or beneficial to certain ends. \textit{Arguments against the use of visual complexity} (i.e. assumptions driving the minimalist mantra) comprise cognitive and emotional (side) effects such as the tendency of complex charts to bewilder, scare, repel, mislead, slow consumption or shut people down, to confuse and overwhelm them, or to hide relevant details and discourage exploration. In short, the guiding minimalist notion is that chart complexity provides no solution, as that's what complex data already does to most of its observers. 

\textit{Arguments for the use of visual complexity}  (i.e. assumptions driving objections to minimalism and to maintain or increase visual complexity) turn the table on many of the pertinent worries and focus on the potential of chart complexity to impress, attract, and to appeal aesthetically, to rise emotional impact, add nuance, allow for polyvocality, promote idiosyncratic interpretation and deeper reflection, to defamiliarize the object of study, promote memorability, convey authority, slow consumption down (as in ``slow analytics''~\cite{bradley2016visualization}), and also to encourage interaction and exploration. Complexity can also be used rhetorically to convince people that a problem really is complicated, because there is no justifiable way to visualize it by simple means \cite{koerth2020why}.

\textcolor{interaction}{\subsection{Interaction}}

\noindent Interactivity is known to aid the comprehension of data across multiple levels of details and perspectives through multiple operations on data or their representation~\cite{munzner2014visualization, yi2007toward}. Thus, it is the main technique to modulate and control upstream and downstream axes of complexity according to the users' developing intent. By enabling the display of data facets across multiple views, interaction can help to reduce or increase complexity dynamically with regards to the visualization and the displayed data, by reducing clutter, or by focusing only on relevant aspects. Most notably, interaction is used to regulate the complexity described with the axes of transformation and visualization. However, despite its purported purpose and potential to effectively solve many complexity challenges, interaction introduces a complexity layer of its own on top of every representational system or device. 

Part of this added complexity is commonly referred to as ``interaction costs''~\cite{lam2008framework}. Such interaction costs can result from the required additional cognitive work in the decision- and sensemaking cycle. For instance, an abundance of options to interact, together with multi-level interaction patterns demand more visual and mental effort for analytical decision-making in contrast to limited interaction options and single-level interfaces. Here, the complexity of the interaction is a trade-off between the exploratory freedom of a user and a lower cognitive load~\cite{lam2008framework}. Similarly, motor effort (e.g., hitting small buttons with a pointer) and physical effort (e.g., distance of pointer movements and quantity of clicks) tend to increase the operational complexity~\cite{lam2008framework}. With this, the complexity of interaction work varies between low efforts for simple interactions (e.g. scrolling) and the steeply increased effort for using advanced and combined interaction techniques (e.g. multi-touch gestures or combination of multiple consecutive interaction patterns). Another cost is, in environments where many visualizations are presumed to be static, signaling the interactive affordances of a given visualization~\cite{boy2015suggested}, or providing alternate methods of communication for those who do not interact with a chart in sufficient detail.

A further layer introduced through interaction is the temporal dimension. While static data representations oftentimes suffer from spatial limitations of page or screen sizes, interaction is able to separate the representation into multiple views with the trade-off that it introduces time as another dimension~\cite{yi2007toward}. Temporal separation also is relevant with regard to directness of an interaction. Keeping temporal, spatial and conceptual separation low can improve directness of interactions~\cite{tominski2020interactive} and by this decrease cognitive complexity.

Also the familiarity and experience of users with an interaction techniques or patterns has influence on the perceived complexity. Training or experience may reduce the demanded mental effort and therefore the perceived complexity. In contrast, when using novel interaction patterns, unexpected behaviour (e.g., non-standard zoom behaviour in maps) of an interface can lead to confusion or frustration for unfamiliar users~\cite{lam2008framework}. 

\smallskip
\noindent\textbf{Uses and abuses:} In the end, interaction complexity is a balancing act between considerations of user freedom, expertise, familiarity, and the goals of an interactive visualization. For instance, scrollytelling pieces are capable of narrating complex topics with complex visual representations: A narrative introduction to machine learning \cite{Yee2018} can be considered as complex regarding encoding and content, while the interaction itself is minimized to scrolling-based  navigation of the information in an author-driven way. 

In contrast, a t-SNE based world map , can arrange countries in form of circles based on similarity metrics~\cite{Rokotyan2019}. The encoding of the visualization itself is rather simple, however, the interfaces offers many options and possibilities to adjust the representation. It allows selection of the countries for detailed information, their re-coloring and re-sizing, in joint with adjusting the weights of the similarity calculations with fine-tuned sliders for over 30 data dimensions. Users are not only provided with a lot of freedom for exploration, but also with an abundance of possibilities to choose. Despite rather simple interface elements for interactions (drop-down menus, sliders, check boxes), the multitude of options and possibilities increases the complexity of the overall visualization. Additionally, the abstractness of t-SNE calculations adds to the complexity, by making results of interactions and new layouts rather unpredictable.

\textcolor{interpretation}{\subsection{Interpretation}}

\noindent A common framing of work from graphical perception that influences visualization design is that interpreting visualization is primarily a matter of \textit{decoding}: that is, taking a data value that has been \textit{encoded} via a visual variable (like height in bar charts) and converting it back into a value through visual estimation. Bertini et al.~\cite{bertini2020shouldn} challenge this conception. For one, the decoding of individual values might be just one of the steps involved in extracting the desired information from a visualization---other tasks might involve statistical judgements around aggregate values, or cross multiple levels of description and summarization~\cite{brehmer2013multi} that involve not just decoding values but performing additional sensemaking. For another, mismatches in visual metaphors, genre, or visual literacies in the intended audience can produce errors or misinterpretations in visualizations even if lower-level perceptual skills are correctly applied~\cite{nobre2024reading}.

We refer to interpretative complexity as the relative ease or difficulty with which a viewer successfully reads and interprets a visualization. This difficulty is based on several sub-components:

\noindent (1) \textit{Visual literacy and chart familiarity}. Unfamiliar designs or genres of charts can be difficult to interpret, and can require active effort to understand~\cite{lee2015people}, and occasionally even explicit designs to ``onboard''~\cite{dhanoa2022process} new audiences. Likewise, dis-congruent or unfamiliar visual metaphors can also impact task performance and chart fluency~\cite{ziemkiewicz2008shaping}.

\noindent (2) \textit{Task and analytical intent}. While ``decoding'' a single value (say, determining the height of a bar in a bar chart and then reading off the corresponding value on the y-axis) is straightforward, analytical goals and insights in charts are rarely at the level of reading off a single value, but can involve \textit{aggregate}~\cite{szafir2016four} tasks, or even statistical judgements and estimations (like visually estimating trends~\cite{correll2017regression}). The notion of intepretational complexity is particularly apposite when considering \textit{insights}. Per Shneiderman~\cite{hullman2019purpose}, ``the purpose of visualization is insight, not pictures'': and the nature of these insights, even those derived from relatively simple displays or data, are inherently ``complex'' and ``deep''~\cite{north2006toward}.

\noindent (3) \textit{Grounding and explanation}. Even if the proper value(s) have been successfully read from the chart, there is still the resulting step of translating those values within the context of use (i.e., what decoded visual features mean in actual interpretative terms), which can vary in complexity. To revisit an example from our introduction, the \textit{meaning} of a difference between two points in a scatterplot is dependent on their axes. If the axes are ``simple'' (say, sales over time), then the interpretation of a point might also be simple (say, a point might be two months in the future, and have twice as many sales, as another). However, axes generated by processes like multidimensional scaling or projection can be difficult to explain~\cite{gleicher2013explainers,faust2018dimreader} or conceptualize, involving combinations of multiple underlying data dimensions in occasionally non-linear geometric space. The common visual form of the scatterplot is identical between these two cases, but the complexities of interpretation are not. We also note that forms of hermeneutical inquiry like close reading~\cite{bares2020close} can involve considerable interpretative effort even if the underlying data are relatively ``simple'': the process of \textit{reading into} a work can introduce interpretative complexity.

\smallskip
\noindent\textbf{Uses and abuses:} As per Bertini et al.~\cite{bertini2020shouldn}, designs that may be more ``complex'' from the perspective of decoding individual values may still have benefits either for supporting the intended tasks of the visualization, better aligning with existing or expected visual metaphors. There are other purported benefits to what would otherwise be seen as ``complex'' interpretative designs: ``beneficial difficulties''~\cite{hullman2011benefitting} that can improve engagement and promote a ``slow analytics''~\cite{bradley2016visualization} or ``slow reading''~\cite{fragapane2022infographics} with deeper and more thoughtful engagement with the material. We note, however, that techniques that superficially appear to simplify data (like projection, clustering, modeling, and sampling) can incur follow-up costs in terms of interpretative complexity. Machine learning might be a quintessential example: a predictive algorithm might condense a high-dimensional space down to a binary ``yes/no'' decision. This decision is trivial to visualize. However, good designs that afford a simple \textit{interpretation} of this decision occupies the expansive field of explainable AI (XAI) research.

\begin{figure*}[t!]
\includegraphics[width=\textwidth, alt={Synthesis}]{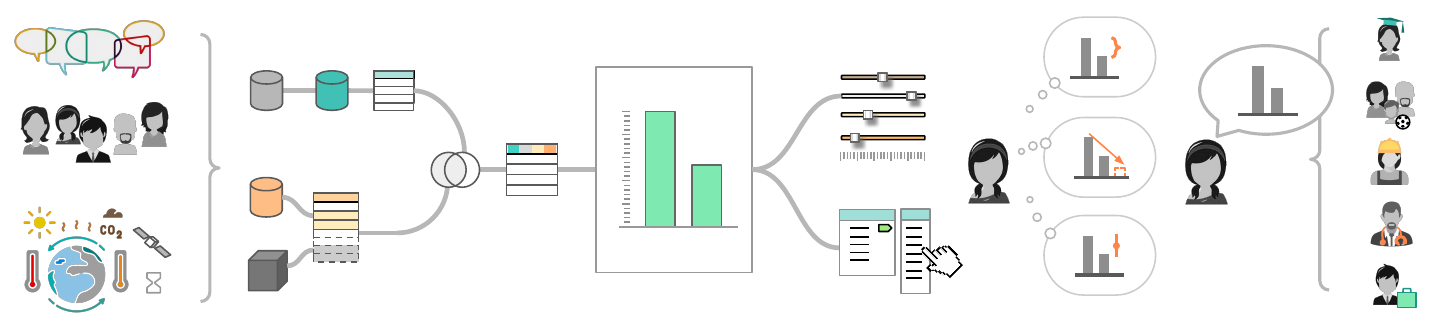}
  \caption{Even a superficially ``simple'' visualization like a bar chart can exhibit entire chains of interconnected complexities, each with their own associated design decisions. These decisions about complexity are associated not only with the data that are eventually used to create the chart, but also how a viewer might interact with the chart, derive insights from the data contained within, and communicate those insights to diverse audiences.}
  \label{fig:bar-iceberg}
\end{figure*}

\textcolor{communication}{\subsection{Communication}}

\noindent Another stage of complexity management comes into play if visualization interpretation should be optimized and actively supported for more than one user group. For one, visualizations do not reveal or disclose themselves, but depend on different types of mediation, onboarding, contextualization, presentation, argumentation, persuasion, or explanation strategies \cite{stoiber2023visualization}. For another, depending on the aims and context of a visualization project, the addressed audience can be complex (heterogeneous, diverse, differentiated) in itself. Visualization designers thus are facing distinct user groups with different levels of prior knowledge, motivation, preferences, intentions, and tasks. As a consequence, the conditions of success multiply and require multiple efforts of communication design.

For visualization projects addressing more than one user group, designers have to consider developing either multiple user-specific i) onboarding and communication programs~\cite{dhanoa2022process} (e.g. with varying depth for lay persons or experts) or ii) multi-user interface designs (i.e. with user-specific combinations of visualization and interaction design). Related design strategies have been discussed, amongst others, with focus on adaptive, personalized, or customizable interfaces \cite{ottley2022adaptive}, progressive disclosure \cite{springer2019progressive}, multi-channel approaches to data visualization \cite{wood2014moving}, design for expert vs. casual users \cite{pousman2007casual} and persona-driven \cite{salminen2022use}, or accessible, inclusive, and universal design \cite{marriott2021inclusive}. 

Connecting back to axis 1 (topic complexity, defined during a project's \textit{initiation} phase), the communication scenarios for some topics are known to be specifically complex, as they are widely present in contemporary culture, politics, and media, or even argued to be of planetary concern, but with different implications for different population subgroups (e.g., climate change). Related knowledge, discourse, and decision scenarios are commonly complex, and full of polarizing epistemological, political, ethical, and interpretative controversies. Under such circumstances, visualization-based communication arguably becomes hyper-complex, requiring designers not only to think about all available measures of inclusive, appealing, and engaging design, but also to coordinate their local project initiatives with the complex communication and engagement strategies of large (inter- and non-) governmental activity programs \cite{windhager2019inconvenient, schuster2024being}.  

\smallskip
\noindent\textbf{Uses and abuses:} The inherent strengths of visualizations to reduce and accessibly convey topic and data complexity for a wide range of audiences provides a main rationale for the whole field \cite{nino2008diagrams}. As such, its applied, communication-supporting branches weave through all fields of society, including science \cite{franconeri2021science}, business \cite{kernbach2015use}, journalism \cite{fu2023more}, and politics \cite{naerland2020political}. Arguably, the discussion of visualization and interpretation complexity is of specific relevance for communication endeavors addressing heterogeneous audiences with a large share of non-expert or casual users, where prior knowledge, literacy and motivation to engage with parsimonious designs vary starkly \cite{pousman2007casual, hung2018affective}. 

Strategies to misuse simplified or distorted representations of data complexity in this context are even-aged with the whole field \cite{cairo2019charts, lisnic2023misleading} and continuously updated \cite{hannah2021conspiracy}. But also the deceptive complexification of adversarial topics (see e.g., Figure 5 in \cite{correll2017black}) is part of the deceptive playbook. As for corrective antidotes against these visualization and complexity ``for villainy''~\cite{mcnutt2021visualization}, the expansion of data and visualization literacy initiatives \cite{camba2022identifying, borner2019data}, visualization provenance standards \cite{ragan2015characterizing}, and critical design strategies~\cite{dork2013critical, d2016feminist} are of the essence. \\

\section{Discussion}
\label{sec:discussion}

We present the discussions of complexity in the previous sections of the paper not to present a complete exploration of the concept of complexity as it might relate to visualization---as previously mentioned, given the polysemic and occasionally ambiguous nature of complexity as a concept, any attempt to be complete is likely doomed to failure. Rather, we present our axes of complexity to highlight two important aspects overlooked in existing debates around complexity:

\noindent (1) To highlight that \textbf{complexity is not an exclusive property of the visual design of a chart}. Even an almost prototypical ``simple'' visualization like a bar chart can be the product of (and result in) a complex web of interconnected complexities (\autoref{fig:bar-iceberg}), akin to the ``referential'' \cite{latour1999circulating} and ``tentacular'' \cite{haraway2018staying} entanglements traced by science-and-technology studies: upstream complexities arising from the long chains of provenance and preparation of data, as well as downstream complexities, resulting from analytical and interpretative tasks, and even from strategies for communication.

\noindent (2) To show how these various kinds of \textbf{complexity are contingent and interconnected}. Complexity that is seemingly reduced at one stage of the process can pop up again, in another guise, in other sections of the pipeline. Decisions about complexity have interconnected upstream and downstream effects on the overall complexity in a visualization when perceived of as not just a visual design, but an entire system of sense-making. Apparent reductions in complexity could, therefore, be more like tradeoffs: reductions of complexity in one area of the sensemaking process that incur corresponding costs in another area: complexity may not be truly reduced, but merely shifted around.

We therefore echo calls from other areas of HCI and design--- to ``recognize''~\cite{andersson2014recognizing}, ``live with''~\cite{norman2016living}, or ``dance with''~\cite{garcia2014enactive} complexity in our own visualization design practices. We likewise echo the call from Akbaba et al.~\cite{akbaba2021manifesto} to remove terms like ``chartjunk'' from our vocabulary as ultimately limiting and inflexible for describing when, how, and why to make use of complexity in designs. We believe, instead, that the concept of a \textit{design material} captures the way that complexity can be \textit{strategically} employed in the design process, just as other concepts like AI can likewise be thought of as design materials~\cite{dove2017ux,holmquist2017intelligence} whose legibility and utility improve as a function of growing a designer's familiarity and expertise with their incorporation into the design process. 

We point to two examples of the strategic use of complexity that arise from our consideration of the pipeline. The first is an extension of the previously mentioned concepts of ``beneficial difficulties'' for visualization~\cite{hullman2011benefitting}, and related ``slow analytics''~\cite{bradley2016visualization}, the seemingly counter-intuitive notion that including additional visual complexity (for the former) or interaction complexity (for the later) can produce benefits in terms of engagement or understanding. If the goal of visualization is merely minimalist efficiency, then these design choices would seem to be nonsensical. But, just as speed bumps and traffic signals are ubiquitous elements of traffic flow design, so too are these strategic uses of complexity ultimately functional. The second example is that of the use of complexity to support interpretative goals beyond the efficient extraction of values. One motivating example is the Poemage project~\cite{mccurdy2015poemage}: the designers of the visualization tool were uncomfortable presenting a visually complex and difficult to untangle visualization of connections in poetry. But the critics, poets, and close readers that used the tool saw value in the resulting ``beautiful mess'' that supported hermeneutical patterns of sensemaking that thrive off of hitherto unseen connections and enrichment rather than reduction.

There are consequences to the current visual-centric view of complexity in visualization. As we discuss in prior sections, many crucial aspects of a data visualization (like data provenance or intended audience) are often overlooked in the way that we present and evaluate visualizations, to our peril. For instance, the numerous ``visualization mirages''~\cite{mcnutt2020surfacing} that can occur where the insight purportedly extracted from a chart does not survive scrutiny, or the ways that visualizations can be mis-applied or misused once they are appropriated by unintended audiences~\cite{lee2021viral}. The job of a visualization designer is not finished merely because they have generated the visually simplest chart that contains all of the data of interest.

\subsection{A Call to Action and Future Work}
To revisit a metaphor from our introduction, just as making proper use of color as a design material in visualization requires knowledge of the human visual system, research on color perception, and potential assistive systems for, e.g., selecting color-blind safe but effective color palettes, so too do we believe that treating complexity as a design material seriously will require considerable future effort of researchers, designers, and practitioners of data visualization. We point to several areas where we see either a lack of existing research, or existing opportunities to translate or apply research from other fields to the specific case of complexity for visualization. In particular:

\noindent (1) New \textbf{metrics and measurements} for complexity, especially non-visual complexities. For viusal complexity, in addition to contentious metrics like Tufte's data-ink ratio~\cite{tufte2014visual}, there are also metrics motivated by psychophysics for concepts like visual clutter~\cite{rosenholtz2007measuring}. What are equivalent metrics for our other axes of complexity? Are there ways of capturing, for instance, that a particular data set is reliant on a highly complex set of underlying data transformations? Or that the intended audience of a visualization might have an diverse and complex array of incompatible analytical tasks or objectives?

\noindent (2) Empirical study of \textbf{intersections and inter-relations} between differing types of complexity. Our supposition that complexity is often not \textit{eliminated} by actions taken to reduce it, but often merely \textit{moved} to other facets of the sense-making process, is just that: a \textit{supposition}, supported by adages from design like Tesler's law~\cite{saffer2010designing} and by point examples from the brainstorming sessions that formed the core of this paper. Further research is needed to establish the borders and constraints of such assertions in visualization design, and, particularly, to assess \textit{where} complexity shifts when it is reduced with respect to one axis.
    
\noindent (3) New designs and techniques for \textbf{usefully surfacing} complexity in visualizations. We lack consistent and widely adopted techniques (or even strategies) for surfacing the ``hidden'' complexities in visualizations, or places for designers to signal where complexity was intentionally reduced (or increased).

\noindent (4) Lastly, new \textbf{pedagogies and frameworks} for teaching and conceptualizing visualization design. Somewhat ironically, we call for a more nuanced and ultimately \textit{complex} view of complexity in the way that we think of visualization design.

\acknowledgments{
 This work was initiated by the Dagstuhl Seminar 23381: ``Visualization and the Humanities: Towards a Shared Research Agenda.'' We also thank Linda Freyberg and Bridget Moynihan for their help in the conceptualization of this paper, as well as Mathieu Jacomy, Fabian Offert, and Menna El-Assady for their contributions to the initial discussion that led to this work.
 }



\bibliographystyle{abbrv-doi-hyperref}

\bibliography{template}

\end{document}